# Knowledge is non-fungible

**César A. Hidalgo**

Center for Collective Learning, ANITI, TSE, IAST, IRIT, University of Toulouse,

Alliance Manchester Business School, University of Manchester

School of Engineering and Applied Science, Harvard University

What would you do if you were asked to "add" knowledge?[*] Would you say that one" plus one knowledge" is two "knowledges"? Less than that? More? Or something in between?

Adding knowledge sounds strange, but it brings to the forefront questions that are as fundamental as they are eclectic. These are questions about the nature of knowledge and about the use of mathematics to model reality. In this chapter I explore the mathematics of adding knowledge starting from what I believe is an overlooked but key observation: the idea that knowledge is non-fungible.

**A brief history of knowledge**

During the XX[th] century our quantitative understanding of knowledge grew thanks to important contributions from psychologists, sociologists, engineers, and economists.

---

[*] Here I am using knowledge as a shorthand for productive knowledge, which is the knowledge that is expressed in a productive activity, from the export of a product, to the creation of a cultural good. This is close to the idea of procedural knowledge, but it also involves collective forms of procedural knowledge (e.g. the ability of a company to build a product is not identical to that of craftsman to produce a craft). Yet, the non-fungibility of knowledge also applies to other forms of knowledge, such as knowledge of facts—descriptive knowledge—or knowledge of things—acquaintance knowledge.

Leon Thurstone—a mechanical engineer turned psychologist—kickstarted the quantitative understanding of knowledge at the dawn of the century by using learning curves to model the growth of knowledge in a typing class involving dozens of students (*1*). Thurstone's work was followed a few decades later by Theodore Wright, an engineer who in the 1930s mapped the learning curves describing cost reductions in aircraft manufacturing (*2*). Eventually, Thurstone and Wright's ideas made it into economics, where scholars such as Leonard Rapping used them to study differences in productivity across shipyards. Armed with econometric models, Rapping was able to show that some increases in productivity were not due to scale or technology, but due to learning within teams (*3*).

By the end of the 20th century knowledge was no longer a vague idea, but a quantifiable concept that had taken center stage in models of endogenous economic growth (*4*–*7*) and in empirical studies (*8*, *9*).

But the quantitative study of knowledge was made possible by some deep qualitative observations. From a modelling perspective, endogenous growth models (e.g. (*4*)) are not particularly complex. They involve a few differential equations. But to motivate these models, economists had to embrace some key concepts, such as the idea that knowledge was a non-rival good. This meant that unlike a capital good, such as a hammer, knowledge could be consumed simultaneously by multiple people. It could be copied without being taken. The consequence of this property of knowledge was profound. It meant that knowledge could grow in per-capita terms in ways that physical capital could not. Knowledge was the essence of economic growth.

Today, scholars agree on a few characteristics of knowledge. Scholars agree that knowledge is non-rival, that it can be tacit or explicit (*10*, *11*), and that organizations and regions often differ in their capacity to absorb it (Cohen and Levinthal's idea of absorptive capacity (*12*)). But there are also some agreed upon puzzles, such as the fact that knowledge has difficulties diffusing, especially for more complex economic activities (*8*, *9*, *13*–*15*). This has motivated scholars to study other characteristics of knowledge, such as its complexity (*13*, *16*, *17*) and the idea that it is non-fungible[†]. In this chapter, I would like to focus on these two ideas, with a particular focus on the non-fungibility of knowledge.

**Knowledge is non-fungible**

Consider swapping a surgeon with a pianist in the middle of an active surgery. Now consider switching the pianist with a surgeon in the middle of a piano concert. Both substitutions will lead to failure, not because surgeons are more or less skilled than pianists, but because the knowledge they have is highly specific. Their knowledge is not interchangeable, or in technical terms, it is non-fungible.[‡]

The non-fungibility of knowledge is the idea that you cannot simply exchange knowledge for knowledge. You cannot change people with different skills (pianist with a surgeon) or documents with different content (music sheet and a patient's hospital records). Of course, this is something that is especially true among activities that are widely different (pianists and surgeons), and less so among related or equivalent activities (pianists and pianists). Still, even among "identical" activities fungibility may be limited if, in the case of the piano, the substitute pianist has not practiced the piece of music being played by the first one. Knowledge is non-fungible, not in a binary sense, but in a continuum. This non-fungibility, however, implies that you cannot simply add knowledge, since knowledge is not a thing, but a constellation of unique "flavors," "letters," or "categories." Like characters in an alphabet or atoms in chemistry. We can speak of knowledge as if it were a thing, and we can even aggregate it in formal models, but that mathematical convenience will

---

[†] While the idea that knowledge is non-fungible is not commonly discussed in these exact terms in the economics literature, it is present implicitly in many combinatorial models that use non-fungible elements (letters in an alphabet, ingredients in a recipe, etc.). For example (*5*, *18*–*22*).
[‡] This is also true for descriptive forms of knowledge. The content of entries in an encyclopedia is not interchangeable. Going back to our doctor and pianist analogy, swapping a patient's hospital chart with a music sheet also brings to light the non-fungibility of descriptive knowledge.

eventually clash with the empirical reality that knowledge is not a thing, but an "alphabet.§"

The non-fungibility of knowledge has profound conceptual and mathematical implications. One of them is that knowledge cannot be simply aggregated. A surgeon plus a pianist are not equal to two surgeons or two pianists. But a pianist plus a guitarist are the beginning of a band.

Another implication of the non-fungibility of knowledge is that it can help explain constraints to knowledge diffusion. Of course, the tacit nature of knowledge can also help explain difficulties in knowledge diffusion, but it is not enough. In a world where knowledge is non-fungible, and some pieces of knowledge are complementary to others, accumulating knowledge involves accumulating letters from an alphabet or pieces from a puzzle. Missing a few pieces can be enough to ruin the transfer of knowledge between two locations, meaning that the non-fungibility of knowledge, and the combinatorial complexity it implies, are characteristics that help explain the limited diffusion of complex knowledge.

The bottom line is that to study knowledge we need to look beyond our traditional mathematical tools. To explore its non-fungible and complex nature, we need to use ideas from matrix algebra that go beyond traditional forms of aggregation. These are techniques used in theoretical physics, machine learning, and biology, in problems that require preserving the identity of the elements involved (e.g. problems involving genes, proteins, books, etc.). These are tools developed to understand systems of organized complexity (*23*), which are defined as systems were the identity of the elements involved and their patterns of interaction cannot be ignored.

---

§ Knowledge is not only unique and non-fungible, but it can exist at multiple scales. It is available in people, teams, organizations, cities, and countries. These agglomerations, however will tend to involve complementary forms of expertise, so teams—or at least the teams that survive and we get to observe—will accumulate letters that are complementary. Guitar players will look for singers and drummers, and actors for filmmakers and screenplay writers. Knowledge may be composed of "letters" or "atoms," but what we observe in the world is a complicated chemistry or "paragraphs" and "sentences."

**If Knowledge was Fungible**

Let's take a step back and consider a world of fungible knowledge. How would it look like?

In that world, knowledge would behave as a single factor. In the alphabet analogy, this would be a world in which the only thing that matters is the number of letters involved in a word. In that "world," the words *dog*, *bet*, and *log* are the same, since they all require three letters.

In an economy where knowledge is fungible patterns of specialization are simple. They can be described as segments in a line. In a world of fungible knowledge predicting the development of new activities is also easy, since it involves entering the activities that are next in line. Countries, cities, and regions move to products that require one more letter and exit few-letter products. In that world, the space of similar products, or "product space" (*24*), is a chain of beads connecting one letter words, to two letter words, to three letter words, and so on (Figure 1).

## Fungible Knowledge

In a world where knowledge is fungible, development involves moving upwards in a linear scale of sophistication.

Using an alphabet analogy, economies upgrade by moving from 1-letter words to 2-letter words and so forth.

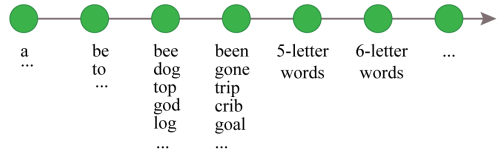

## Non-Fungible Knowledge

In a world where knowledge is non-fungible the patterns of similarity between activities is constrained by overlaps in knowledge.

Using an alphabet analogy, economies upgrade by moving into words that share some letters.

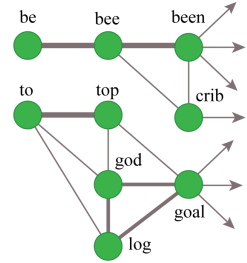

## Networks of Similar Products, Technologies, Industries, etc.

Networks of similar activities estimated using data on products, industries, occupations, technologies, etc. exhibit an intricate structure, as it expected from the idea of non-fungible knowledge. There is also ample work documenting the fact that economies develop by following the paths described in these networks, which is further evidence that knowledge is non-fungible.

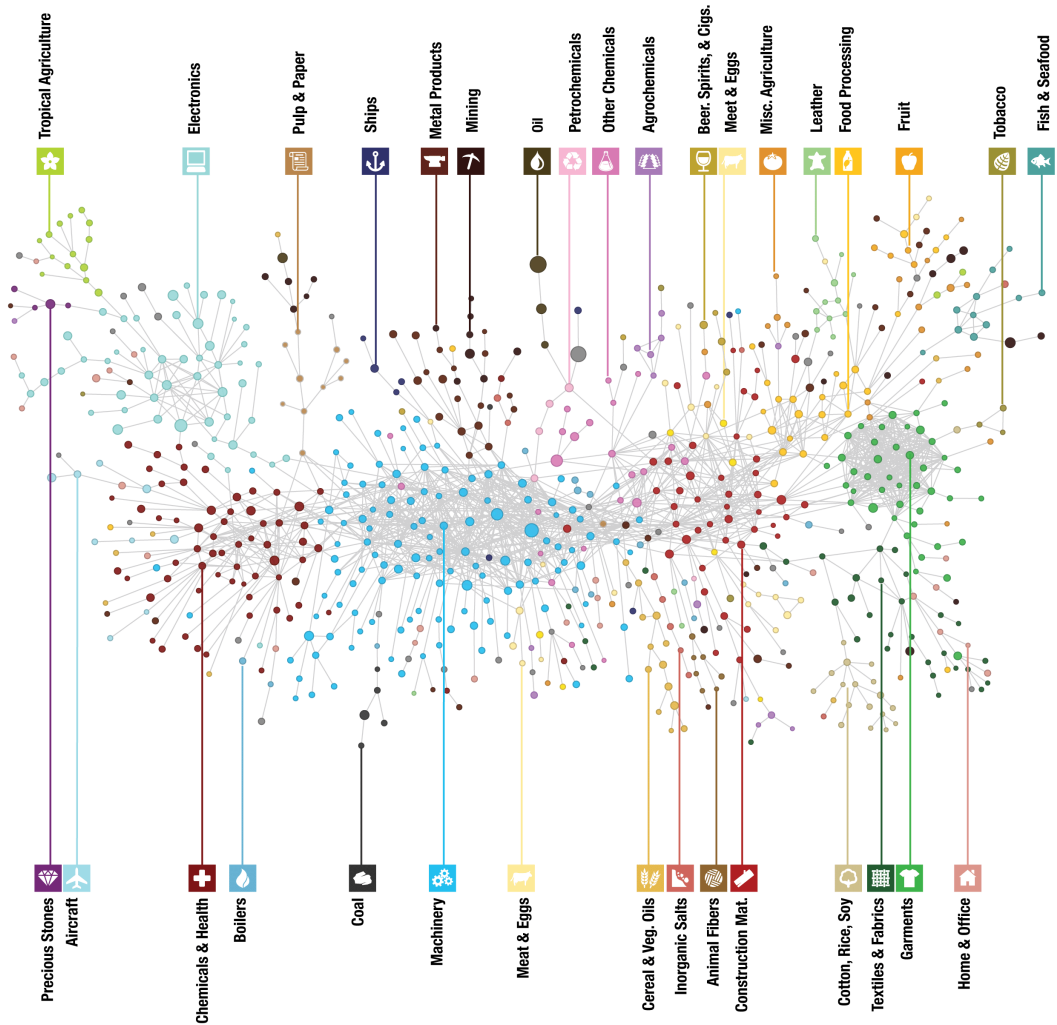

Figure 1. Networks of similar activities in world of fungible and non-fungible knowledge.

But since knowledge is non-fungible, economies evolve through circuitous path dependent dynamics. They do not simply move from three to four letter words, but from three to four letter words that reuse some of the letters they have (Figure 1). In a world where knowledge is non-fungible, economies do not jump from *log* to *camp*, but from *log* to *goal*. The result is an intricate structure, in which patterns of development are constrained by the cognitive relatedness of activities, and where two "equally" knowledgeable economies may face diverging paths. The intricate structures observed in networks of similar products, industries, technologies, and occupations, are a smoking gun evidence of this idea. They are also the tip of the iceberg inviting us to consider the use of matrix algebra methods to answer our original question: how should we add knowledge?

**Eigenvectors all the way to the bottom**

Networks of similar activities help us incorporate the non-fungibility of knowledge by providing a means to construct indicators of the availability of knowledge that are specific to pairs of locations and activities. This is the idea behind the *relatedness density* indicator introduced in 2007 (*24*) and used extensively since in the economic geography literature (*15, 25–30*).

But these networks also have clusters of similar activities that represent overlaps in knowledge, such as the "green" garments cluster on the right-hand side of the product space (Figure 1). This cluster structure tells us that, once we know that a country is able to produce a few types of garment (e.g. shirts, coats, blouses, etc.), we can infer that they have some of the knowledge needed to produce other types of garments. So, the information about the availability of knowledge provided by different products in this cluster is relatively redundant.

This means that these networks, which can be also represented as matrices, can be described by an "alphabet" that is smaller than the number of activities involved, implying that the use of dimensionality reduction techniques on matrices connecting similar activities or locations is a way to measure non-fungible knowledge.

The good news is that matrix algebra provides a solution to this exact problem, since it comes with its own way to create lower dimensional representations of matrices. This is the idea of eigenvectors, a core principle in matrix algebra that helps uncover the natural alphabet of each matrix.

So, what are eigenvectors? How do they work? And how can we apply them to matrices summarizing the geography of economic activities?

Technically, for a square matrix $\boldsymbol{M}$, the eigenvectors $\vec{e}$ are the vectors that, when multiplied by that matrix remain the same except for a constant $\lambda$. That is, they are the solution to:

$$\boldsymbol{M}\vec{e} = \lambda \vec{e}$$

This may seem trivial, but it makes eigenvectors extremely useful, since they are the key to enormously simplifying matrix multiplication. In principle, to multiply a matrix $\boldsymbol{M}$ with a column vector $\vec{v}$ ($\boldsymbol{M}\vec{v}$), we must take the internal product of each row in the matrix and the vector. For large matrices, this can quickly grow to thousands of operations. But if we write that vector as a linear combination of the eigenvectors of the matrix $\boldsymbol{M}$ (as $\vec{v} = a_1\vec{e_1} + a_2\vec{e_2} + \cdots$), we only need to multiply each one of these eigenvectors by its eigenvalue ($\lambda$). This makes matrix multiplication trivial, since it transforms what is a tedious process involving a larger number of operations into multiplying vectors by a constant ($\boldsymbol{M}\vec{v} = a_1\lambda_1\vec{e_1} + a_2\lambda_2\vec{e_2} + \cdots$).

Eigenvectors are a profound concept since they represent the natural alphabet of a matrix. An alphabet given by its eigenvectors $\vec{e}$ with the importance of each given by its respective eigenvalue $\lambda$.

Not surprisingly, eigenvectors (and eigenfunctions more generally) are key concepts in classical and modern physics. The vibrations of a string in a guitar are written in an alphabet given by the discrete solutions to the wave equation (the "eigenvectors" of the wave equation). The frequencies of these vibrations are the eigenvalues. Similarly, the energy levels of an electron in an atom are eigenvalues, and the orbitals, eigenvectors. In modern physics it is eigenvectors all the way to the bottom.

These matrix algebra techniques opened a key epistemological door in the natural sciences. A door to a world in which nature dictates its own alphabet. Before these methods, matter was believed to be made of earth, water, wind, and fire. But eigenvectors and eigenvalues allowed scientists to defy these assumptions by providing a way to interrogate nature using theories that did not assume the nature of factors a-priori, but could learn them a-posteriori. These same ideas, however, can also be applied to economic geography, since they are general enough to discover all types of alphabets, from those of musical notes and orbitals to those of knowledge. In fact, as we will see, this intuition is what lies behind recent ideas on how to add non-fungible knowledge (*31*).

**How should we add knowledge?**

The basic intuition we need to add knowledge is to assume that the activities present, produced, or exported from a location carry key information about the knowledge present in that location. Similarly, the locations where an activity is present can tell us about the knowledge needed to perform each activity. This brings us to a "circular logic" that, as we will see, is also a central idea when it comes to eigenvectors. In simple terms, what we are saying is that cities like San Francisco and Boston are knowledge intense if they are home to knowledge intense activities like biotech and machine learning. In turn, activities like biotech and machine learning are likely to be knowledge intense because they are produced in cities like Boston or San Francisco.

Formally, we let the knowledge $K$ of a location $c$ (e.g. country or city) be $K_c$ and the knowledge $K$ of an activity $p$ (e.g. product or industry) be $K_p$. Also, we let $M_{cp}$ be a matrix summarizing the activities ($p$) present in each location ($c$). Following this notation, the problem of "adding knowledge" is that of solving the following system of equations:

(i) The knowledge of a location ($K_c$) is a function ($f$) of the knowledge ($K_p$) of the activities present in it ($M_{cp}$), and

(ii) The knowledge of an activity ($K_p$) is a function ($g$) of the knowledge ($K_c$) of the places were that activity is present ($M_{cp}$).

This argument is equivalent to solving:

$$K_c = f(M_{cp}, K_p), \quad (1)$$
$$K_p = g(M_{cp}, K_c), \quad (2)$$

Where $f$ and $g$ are functions to be determined.

While this map is quite general, it already rules out some key measures, such as market concentration indexes (e.g. Shannon's information entropy or the Herfindahl-Hirschman index). Measures of concentration fail to couple activities and locations (they only consider shares of activities across locations, but treat all activities as equal).

Equations (1) and (2) can be transformed into two self-consistent equations of the form:

$$K_c = f\left(M_{cp}, g(M_{cp}, K_c)\right), \quad (3)$$
$$\quad (4)$$

$$K_p = g\left(M_{cp}, f(M_{cp}, K_p)\right),$$

Which, for simple forms of *f* and *g* can be reduced—or approximated—by matrix equations of the form:

$$\widetilde{M}_{cc'} K_{c'} = \lambda K_c \quad (5)$$
$$\widetilde{M}_{pp'} K_{p'} = \lambda K_p \quad (6)$$

These equations should look familiar, since they are equations for the eigenvalues and eigenvectors of $\widetilde{M}_{cc'}$ and $\widetilde{M}_{pp'}$: matrices connecting similar locations or similar activities. These equations imply that measures of knowledge ($K$), obtained only form assuming that knowledge is expressed in the geography of activities, can be recovered from the eigenvectors of these square matrices. These eigenvectors are the "alphabet" of the matrix, and the solution to our original problem. These equations also imply that metrics of the knowledge intensity of economies, or of the activities present in them, are respectively, the eigenvectors of matrices of similarity among economies (e.g. countries, cities, regions) ($\widetilde{M_{cc}}$) or activities (e.g. products) ($\widetilde{M_{pp}}$) (*24, 32*). This last point is also key to satisfy a desirable axiomatic property of a measure of knowledge: the notion that economies with similar productive structures should have similar values in that measure.

Now, before we can bring the theory to the data, we need to do one more thing: explore the basic shapes of the matrix $\widetilde{M}_{cc'}$. Here we consider two illustrative cases: *extensive* and *intensive* knowledge.**

Extensive and intensive variables are a key concept in statistical mechanics. Extensive variables are those that scale with the size of a system. Intensive variables do not. Volume, population, and GDP are extensive variables. Temperature, pressure, and GDP per capita are intensive variables.

---

** In principle we can consider many more forms, such as forms that combine extensive and intensive variables, or even forms with reciprocals. In practice, many of these forms provide very similar solutions.

Using an extensive assumption is simple. We just add knowledge across activities, no matter if they overlap or not. In this case *f* and *g* are just sums and the system to solve becomes:

$$K_c = \sum_p M_{cp} K_p$$

$$K_p = \sum_c M_{cp} K_c$$

meaning that we need to find the eigenvectors of the matrix:

$$\widetilde{M}_{cc'} = \sum_p M_{cp} M_{c'p},$$

which is simply $M$ times its transpose ($\widetilde{M}_{cc'}$ is simply the number of activities common to two economies).

The intensive assumption is obtained by averaging instead of adding. A property of this assumption is that our estimate of the knowledge available in a location grows only when an activity is above a location's current average. If we define diversity ($M_c$) as the number of activities in a location ($M_c = \sum_p M_{cp}$) and ubiquity ($M_p$) as the number of locations where an activity is present ($M_p = \sum_c M_{cp}$), then,

$$K_c = \frac{1}{M_c} \sum_p M_{cp} K_p$$

$$K_p = \frac{1}{M_p} \sum_c M_{cp} K_c,$$

and $\widetilde{M}_{cc'}$ takes the form:

$$\widetilde{M}_{cc'} = \frac{1}{M_c} \sum_p \frac{M_{cp} M_{c'p}}{M_p}$$

Notice that for the intensive case $\widetilde{M}_{cc'}$ is a stochastic matrix, meaning that each row adds up to one (sum over $c'$ and everything cancels). This implies that the first eigenvector of this matrix is always an eigenvector of 1s and the second eigenvector is the first one to contain a non-negligible variance. As we will see in the next section, this is a key feature to help avoid measures of knowledge that are too biased towards the size or diversity of an economy. In both cases, we can estimate the eigenvectors by solving:

$$\det(\widetilde{M}_{cc'} - \lambda I) = 0,$$

to find the eigenvalue $\lambda$, and then find the eigenvectors by solving[††]

$$(\widetilde{M}_{cc'} - \lambda) K = 0.$$

Now that we have a worked-out theory, we can bring the data to the model to answer our original question: how should we add knowledge?

**Bringing the theory to the data**

So far, we have talked about adding knowledge in theory. But in practice, data on the geography of economic activities is very heterogenous. This brings in some challenges that require us to work on the data before we can apply the theory.

Consider international trade data. It compares countries as big as China, with more than USD 2 trillion in exports, with Vanuatu, a pacific island with less than USD 200M in exports (a 10,000x difference). A similar unevenness in the size of the units of observation is observed for products (e.g. Crude Oil vs Felt Hats). The same observation can be made for data on cities and for activities as varied as technologies, industries, and occupations. This means that matrices summarizing the location of economic activities do not have rows and columns that can be readily compared. So before bringing the theory to the data, we need to make these units of observation more readily comparable.

---

[††] Although in practice, it is common to use a numerical software (e.g. Matlab).

This can be achieved through a sequence of normalizations and data cleaning procedures. First, we start by cutting the "left tail" of the observations so we can consider only locations and activities that are larger than a certain size (*33*) (e.g. exclude small nations such as Tuvalu, Vanuatu, etc.). Still, the remaining units of observation can be highly uneven (e.g. Uruguay and China), so there are additional steps that we need to take to make them more comparable.

The next step is to transform these matrices into matrices of specialization by calculating the location quotient (LQ) or revealed comparative advantage (RCA). This means to simply normalize matrices by the sum of their rows and columns, which in a probabilistic interpretation, is to take the ratio between the observed and expected value. For a matrix of output $X_{cp}$ (e.g. exports by country and product, patents by city and technology, etc.) we estimate the specialization matrix $R_{cp}$ using:

$$R_{cp} = \frac{X_{cp} X}{X_c X_p}$$

Where once again muted indexes have been added (e.g. $X_c = \sum_p X_{cp}$).

$R_{cp}$ provides a measure of specialization that is more readily comparable than $X_{cp}$, but it still has some problems. On the one hand, values of $R_{cp}>1$ tell us that a location produces more output in an activity than what is expected from its total output ($X_c$) and the output on that activity ($X_p$). That is good. But on the other hand, $R_{cp}$ values can be very large for small countries and activities (small $X_c$ and small $X_p$ in the denominator) and cannot be too large for large countries and activities (large $X_c$ and large $X_p$ in the denominator). So, the variance of $R_{cp}$ is biased. It is larger for smaller economies and activities. This means that the values in the matrix $R$ are still not readily comparable, since a diversified economy, like that of Germany, will rarely have values of $R_{cp}$ larger than 3, whereas smaller economies, like that of Senegal, can easily get values in the 100s for the few products they specialize in. To mitigate this bias, we use one additional normalization. We transform $R_{cp}$ into a binary matrix, $M_{cp}$, which is simply equal to 1 when $R_{cp}$ is larger or equal to one and zero otherwise.

Now that we finally have a matrix with units of observation that are relatively comparable, we can bring the theory to the data.

Figure 2 uses trade data for the year 2020 to compare the first and second eigenvectors of the extensive definition, the second eigenvector of the intensive definition (known as the Economic Complexity Index or ECI), and a simple measure of diversity ($M_c$). Figure 3 does the same using data on payroll by industry for cities (MSAs) in the United States.

We can easily see that the first eigenvector of the extensive definition doesn't do much more than count the number of activities connected to a location. It is highly correlated with diversity ($R^2$=96.6%). This is a problem because we need a measure that goes beyond a simple measure of size. We are looking for a measure that can tell us about the knowledge in an economy that is nevertheless able to highlight small yet sophisticated economies, such as Finland, Singapore, and Taiwan, or put the economies of Boston and Silicon Valley above those of larger metro areas. This is a behavior that we start to see on the second eigenvector of the extensive metric (which is perpendicular to the first), but that is very clear on the second eigenvector of the intensive metric (the Economic Complexity Index or ECI). Remember that the first eigenvector of the intensive metrics is a vector of 1s by construction, so it is unimportant. The intensive metric (ECI) is built on a matrix that forces the units of observation to be comparable (a stochastic matrix where each row adds up to one). This means the rows for China and Uruguay, or those for Boston and Seymour Indiana, "weigh" the same. The difference captured by the vector, therefore, must be a difference in how that weight is distributed along the vector, a difference in structure rather than size.

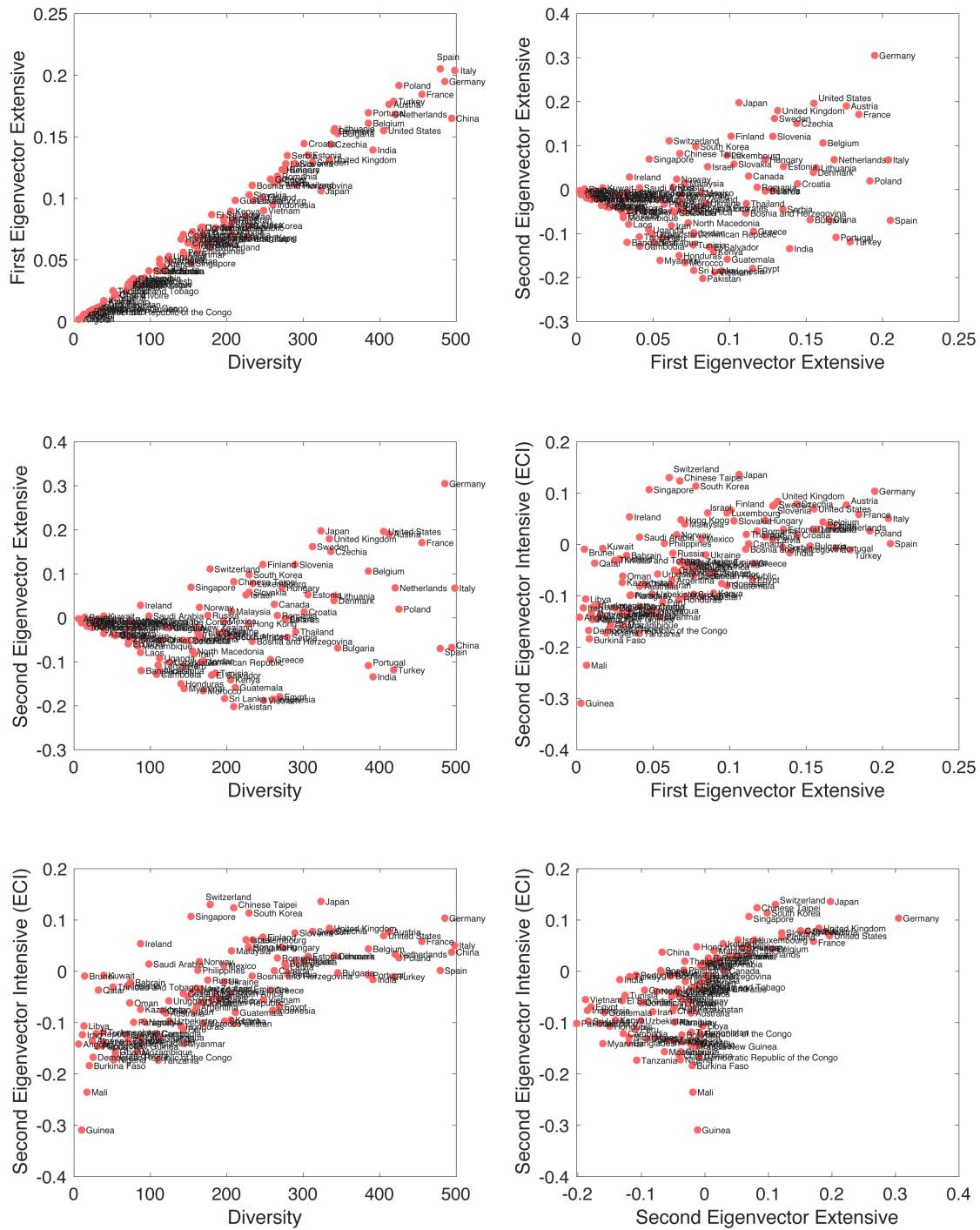

Figure 2. Comparison between eigenvectors of extensive and intensive metrics of knowledge estimated using 2020 international trade data (from BACI/OEC.world)

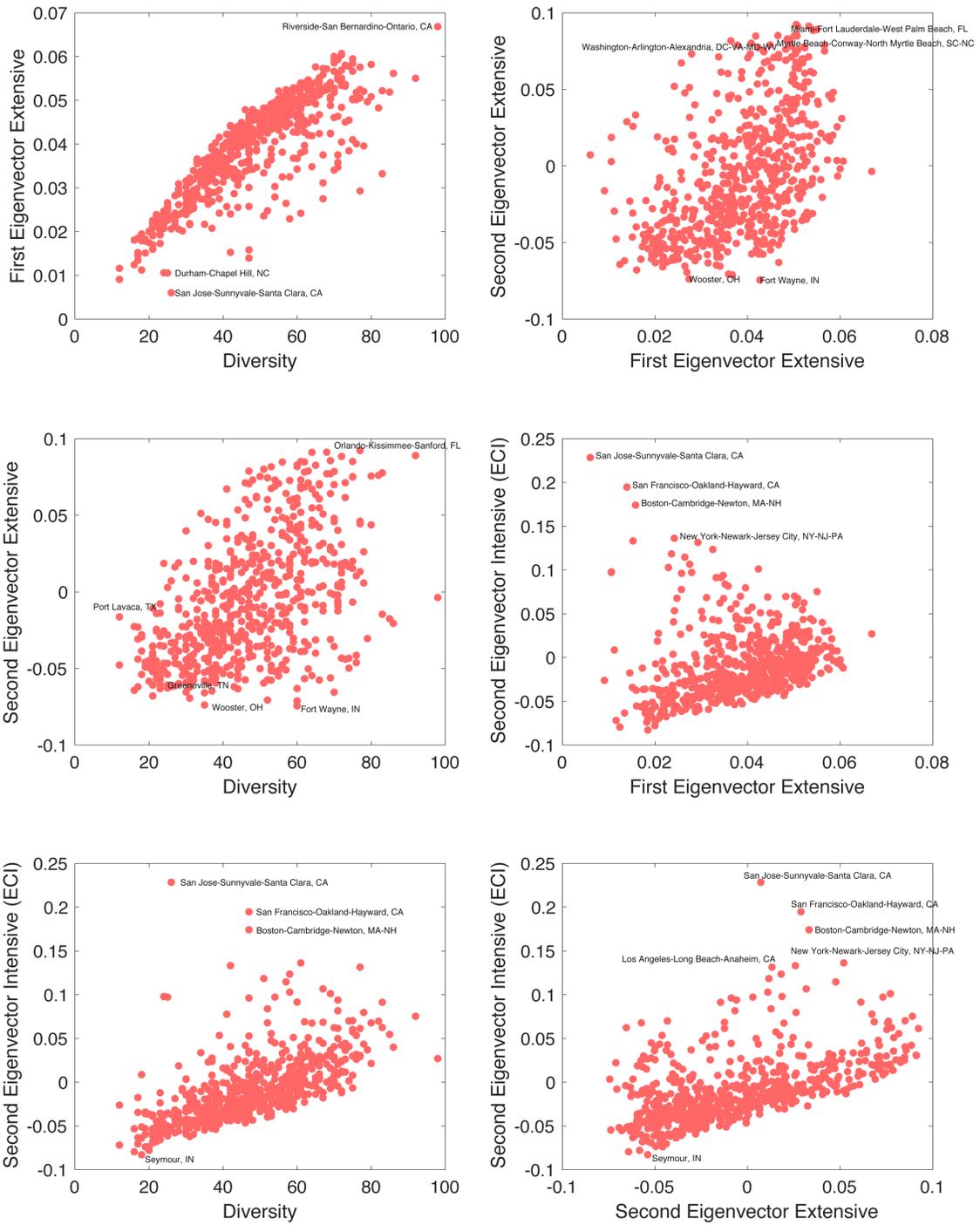

Figure 3. Comparison between eigenvectors of extensive and intensive metrics of knowledge estimated using 2016 payroll by industry and MSA data from the US County Census Business Patterns.

This is why this intensive metric provides a good answer for our original question: "how to add knowledge?" In a system where units of observation are heterogenous, the first eigenvector is obvious, it is just a measure of size. It is the second eigenvector the one containing the non-trivial part of the answer. The one that captures the structure that emerges from the non-fungible nature of knowledge. It is this second eigenvector, the one that has been shown repeatedly to explain variations in economic growth (*16*, *33–41*), inequality (*42–48*), and emissions (*49–56*). There is something about the use matrix algebra techniques to "add knowledge."

**From scales to chords**

We started this essay with a simple but provocative question: What would you do if you were asked to "add" knowledge? Our answer began with a simple but important observation: the idea that knowledge was non-fungible. This simple idea pushed us to reject traditional forms of aggregations and embrace matrix algebra techniques that preserve the identity of the elements involved. We found evidence of the non-fungibility of knowledge in the literature on relatedness and then ventured through the world of eigenvectors to learn the "alphabet" of the geography of knowledge. But bringing the data to the theory was challenging because of the inherent heterogeneity in the units of observation of geographic data. Nevertheless, we were able to provide a detailed path explaining the rationale of many of the decisions that inform modern metrics of non-fungible knowledge, such as the economic complexity index, which can capture information about the knowledge intensity of economies thanks to three key normalizations: the normalization used to define the specialization matrix $R$, the binarization used to mitigate the variance bias of $R$, and the normalization coming from the intensive definition which helps control for differences in size and/or diversity that remain even after the first two normalizations.

But these advances are still narrow in that they involve single matrices. Unfortunately, reality is more complex, meaning that the geography of knowledge is expressed across a variety of activities, not independently, but simultaneously. Eigenvectors are musical notes but the world is made of scales and chords. To get a more complete picture of the geography of knowledge we must complemented trade data with data on the geography of patentable technologies, value added by industry, and scientific publications, to name a few. While there have been efforts to look at economic complexity using data from different sources, no effort yet has looked at multiple sources in combination. This heralds an opportunity for the field. An opportunity to study how to add knowledge across multiple scales and dimensions.

**References**

1. L. L. Thurstone, thesis, Psychological review Co., Princeton, N.J. (1919).

2. T. P. Wright, Factors Affecting the Cost of Airplanes. *Journal of the Aeronautical Sciences*. **3**, 122–128 (1936).

3. L. Rapping, Learning and World War II Production Functions. *The Review of Economics and Statistics*. **47**, 81–86 (1965).

4. P. M. Romer, Endogenous Technological Change. *Journal of Political Economy*. **98**, S71–S102 (1990).

5. M. L. Weitzman, Recombinant Growth. *Q J Econ*. **113**, 331–360 (1998).

6. P. Aghion, P. Howitt, A Model of Growth Through Creative Destruction. *Econometrica: Journal of the Econometric Society*, 323–351 (1992).

7. K. J. Arrow, in *Readings in the Theory of Growth: a selection of papers from the Review of Economic Studies*, F. H. Hahn, Ed. (Palgrave Macmillan UK, London, 1971; https://doi.org/10.1007/978-1-349-15430-2_11), pp. 131–149.

8. A. B. Jaffe, M. Trajtenberg, R. Henderson, Geographic Localization of Knowledge Spillovers as Evidenced by Patent Citations. *Q J Econ*. **108**, 577–598 (1993).

9. M. P. Feldman, D. B. Audretsch, Innovation in cities:: Science-based diversity, specialization and localized competition. *European Economic Review*. **43**, 409–429 (1999).

10. H. Collins, *Tacit and Explicit Knowledge* (University of Chicago Press, 2010).

11. H. M. Collins, The TEA Set: Tacit Knowledge and Scientific Networks. *Science Studies*. **4**, 165–185 (1974).

12. W. M. Cohen, D. A. Levinthal, Absorptive Capacity: A New Perspective on Learning and Innovation. *Administrative Science Quarterly*. **35**, 128–152 (1990).

13. P.-A. Balland, D. Rigby, The Geography of Complex Knowledge. *Economic Geography*. **93**, 1–23 (2017).

14. P.-A. Balland, C. Jara-Figueroa, S. G. Petralia, M. P. A. Steijn, D. L. Rigby, C. A. Hidalgo, Complex economic activities concentrate in large cities. *Nat Hum Behav*, 1–7 (2020).

15. B. Jun, A. Alshamsi, J. Gao, C. A. Hidalgo, Bilateral relatedness: knowledge diffusion and the evolution of bilateral trade. *Journal of Evolutionary Economics*, 1–31 (2019).

16. C. A. Hidalgo, R. Hausmann, The building blocks of economic complexity. *PNAS*. **106**, 10570–10575 (2009).

17. L. Fleming, O. Sorenson, Technology as a complex adaptive system: evidence from patent data. *Research Policy*. **30**, 1019–1039 (2001).

18. R. Hausmann, C. A. Hidalgo, The network structure of economic output. *Journal of Economic Growth*, 1–34 (2011).

19. F. Tria, V. Loreto, V. D. P. Servedio, S. H. Strogatz, The dynamics of correlated novelties. *Scientific reports*. **4**, 5890 (2014).

20. T. M. A. Fink, M. Reeves, R. Palma, R. S. Farr, Serendipity and strategy in rapid innovation. *Nature Communications*. **8**, 2002 (2017).

21. T. M. A. Fink, M. Reeves, How much can we influence the rate of innovation? *Science Advances*. **5**, eaat6107 (2019).

22. C. A. Hidalgo, Economic complexity: From useless to keystone. *Nature Physics*. **14**, 9–10 (2018).

23. W. Weaver, Science and Complexity. *Am Sci*. **36**, 536 (1948).

24. C. A. Hidalgo, B. Klinger, A.-L. Barabási, R. Hausmann, The Product Space Conditions the Development of Nations. *Science*. **317**, 482–487 (2007).

25. Hidalgo, P.-A. Balland, R. Boschma, M. Delgado, M. Feldman, K. Frenken, E. Glaeser, C. He, D. F. Kogler, A. Morrison, F. Neffke, D. Rigby, S. Stern, S. Zheng, S.

Zhu, in *Unifying Themes in Complex Systems IX*, A. J. Morales, C. Gershenson, D. Braha, A. A. Minai, Y. Bar-Yam, Eds. (Springer International Publishing, 2018), *Springer Proceedings in Complexity*, pp. 451–457.

26. R. Boschma, A. Minondo, M. Navarro, The Emergence of New Industries at the Regional Level in Spain: A Proximity Approach Based on Product Relatedness. *Economic Geography*. **89**, 29–51 (2013).

27. D. F. Kogler, D. L. Rigby, I. Tucker, Mapping Knowledge Space and Technological Relatedness in US Cities. *European Planning Studies*. **21**, 1374–1391 (2013).

28. C. Jara-Figueroa, B. Jun, E. L. Glaeser, C. A. Hidalgo, The role of industry-specific, occupation-specific, and location-specific knowledge in the growth and survival of new firms. *PNAS*. **115**, 12646–12653 (2018).

29. J. Gao, B. Jun, A. 'Sandy' Pentland, T. Zhou, C. A. Hidalgo, Spillovers across industries and regions in China's regional economic diversification. *Regional Studies*, 1–16 (2021).

30. R. Boschma, P.-A. Balland, D. F. Kogler, Relatedness and technological change in cities: the rise and fall of technological knowledge in US metropolitan areas from 1981 to 2010. *Ind Corp Change*. **24**, 223–250 (2015).

31. C. A. Hidalgo, Economic complexity theory and applications. *Nature Reviews Physics*, 1–22 (2021).

32. C. Sciarra, G. Chiarotti, L. Ridolfi, F. Laio, Reconciling contrasting views on economic complexity. *Nature communications*. **11**, 1–10 (2020).

33. R. Hausmann, C. A. Hidalgo, S. Bustos, M. Coscia, A. Simoes, M. A. Yildirim, *The atlas of economic complexity: Mapping paths to prosperity* (MIT Press, 2014).

34. J. C. Chávez, M. T. Mosqueda, M. Gómez-Zaldívar, Economic complexity and regional growth performance: Evidence from the Mexican Economy. *Review of Regional Studies*. **47**, 201–219 (2017).

35. G. Domini, "Patterns of specialisation and economic complexity through the lens of universal exhibitions, 1855-1900" (LEM Working Paper Series, 2019).

36. P. Koch, Economic Complexity and Growth: Can value-added exports better explain the link? *Economics Letters*. **198**, 109682 (2021).

37. A. Lo Turco, D. Maggioni, The knowledge and skill content of production complexity. *Research Policy*, 104059 (2020).

38. G. Ourens, Can the Method of Reflections help predict future growth? *Documento de Trabajo/FCS-DE; 17/12* (2012).

39. V. Stojkoski, Z. Utkovski, L. Kocarev, The Impact of Services on Economic Complexity: Service Sophistication as Route for Economic Growth. *PLOS ONE*. **11**, e0161633 (2016).

40. V. Stojkoski, L. Kocarev, The relationship between growth and economic complexity: evidence from Southeastern and Central Europe (2017).

41. B. Doğan, S. Ghosh, I. Shahzadi, D. Balsalobre-Lorente, C. P. Nguyen, The relevance of economic complexity and economic globalization as determinants of energy demand for different stages of development. *Renewable Energy* (2022), doi:10.1016/j.renene.2022.03.117.

42. D. Hartmann, M. R. Guevara, C. Jara-Figueroa, M. Aristarán, C. A. Hidalgo, Linking Economic Complexity, Institutions, and Income Inequality. *World Development*. **93**, 75–93 (2017).

43. L. K. Chu, D. P. Hoang, How does economic complexity influence income inequality? New evidence from international data. *Economic Analysis and Policy*. **68**, 44–57 (2020).

44. M. Ben Saâd, G. Assoumou-Ella, Economic Complexity and Gender Inequality in Education: An Empirical Study. *Economics Bulletin*. **39**, 321–334 (2019).

45. F. Fawaz, M. Rahnama-Moghadamm, Spatial dependence of global income inequality: The role of economic complexity. *The International Trade Journal*. **33**, 542–554 (2019).

46. R. Barza, C. Jara-Figueroa, C. Hidalgo, M. Viarengo, Knowledge Intensity and Gender Wage Gaps: Evidence from Linked Employer-Employee Data (2020).

47. R. Basile, G. Cicerone, Economic complexity and productivity polarization: Evidence from Italian provinces. *German Economic Review* (2022), doi:10.1515/ger-2021-0070.

48. A. Sbardella, E. Pugliese, L. Pietronero, Economic development and wage inequality: A complex system analysis. *PloS one*. **12** (2017).

49. M. Can, G. Gozgor, The impact of economic complexity on carbon emissions: evidence from France. *Environmental Science and Pollution Research*. **24**, 16364–16370 (2017).

50. G. Dordmond, H. C. de Oliveira, I. R. Silva, J. Swart, The complexity of Green job creation: An Analysis of green job development in Brazil. *Environment, Development and Sustainability*, 1–24 (2020).

51. L. Fraccascia, I. Giannoccaro, V. Albino, Green product development: What does the country product space imply? *Journal of cleaner production*. **170**, 1076–1088 (2018).

52. R. Hamwey, H. Pacini, L. Assunção, Mapping green product spaces of nations. *The Journal of Environment & Development*. **22**, 155–168 (2013).

53. A. Lapatinas, A. Garas, E. Boleti, A. Kyriakou, Economic complexity and environmental performance: Evidence from a world sample (2019).

54. P. Mealy, A. Teytelboym, Economic complexity and the green economy. *Research Policy*, 103948 (2020).

55. O. Neagu, The Link between Economic Complexity and Carbon Emissions in the European Union Countries: A Model Based on the Environmental Kuznets Curve (EKC) Approach. *Sustainability*. **11**, 4753 (2019).

56. J. P. Romero, C. Gramkow, Economic complexity and greenhouse gas emissions. *World Development*. **139**, 105317 (2021).